\documentstyle[aps,epsf]{revtex}  

\newcommand{\KZ}{\ensuremath{K^o}}
\newcommand{\DZ}{\ensuremath{D^o}}
\newcommand{\BZ}{\ensuremath{B^o}}
\newcommand{\KZB}{\ensuremath{\overline{K}^o}}
\newcommand{\DZB}{\ensuremath{\overline{D}^o}}
\newcommand{\BZB}{\ensuremath{\overline{B}^o}}
\newcommand{\Dz}{\ensuremath{D^o}}
\newcommand{\Dzb}{\ensuremath{\overline{D}^o}}
\newcommand{\DDb}{\ensuremath{D^o - \overline{D}^o}}
\newcommand{\Dstp}{\ensuremath{D^{*+}}}
\newcommand{\Dst}{\ensuremath{D^{*}}}
\newcommand{\dM}{\ensuremath{\delta m}}
\newcommand{\MD}{\ensuremath{M_{K\pi}}}
\newcommand{\nfb}{\ensuremath{5.6\ {\rm fb}^{-1}}}
\newcommand{\mtws}{\ensuremath{\langle\tau_{\rm WS}\rangle}}
\newcommand{\dstdecay}{\mbox{\ensuremath{\Dstp \rightarrow \Dz \pi^+}}}
\newcommand{\rsdecay}{\mbox{\ensuremath{\Dz \rightarrow K^- \pi^+}}}
\newcommand{\rsbdecay}{\mbox{\ensuremath{\Dzb \rightarrow K^+ \pi^-}}}
\newcommand{\wsdecay}{\mbox{\ensuremath{\Dz \rightarrow K^+ \pi^-}}}
\newcommand{\wsdcsd}{\mbox{\ensuremath{\Dz \stackrel{\rm DCSD}{\rightarrow} K^+ \pi^-}}}
\newcommand{\tbdecay}{\mbox{\ensuremath{\Dz \rightarrow \pi^+ \pi^- \pi^0}}}
\newcommand{\nsdecay}{\mbox{\ensuremath{\Dz \rightarrow K \pi}}}
\newcommand{\mdecay}{\mbox{
\ensuremath{\Dz \rightarrow \Dzb \rightarrow K^+ \pi^-}}}
\newcommand{\pmixinglim}{\ensuremath{1.1 \times 10^{-2}}}
\newcommand{\DG}{\ensuremath{\Delta \Gamma}}
\newcommand{\DM}{\ensuremath{\Delta M}}
\newcommand{\prate}{\ensuremath{0.31 \pm 0.09 \pm 0.07}}
\newcommand{\rslife}{\ensuremath{0.97 \pm 0.02}}
\newcommand{\wslife}{\ensuremath{0.6 \pm 0.2}}
\newcommand{\siglife}{\ensuremath{0.65 \pm 0.40}}
\newcommand{\sigfrac}{\ensuremath{17.3 \pm 5.4}}
\newcommand{\dzfrac}{\ensuremath{8.9 \pm 0.8}}
\newcommand{\othfrac}{\ensuremath{5.8 \pm 0.1}}
\newcommand{\dzct}{\ensuremath{c\tau_{\Dz}}}

\newcommand{\Rdcsd}{\ensuremath{R_{\rm DCSD}}}
\newcommand{\Rmix}{\ensuremath{R_{\rm Mix}}}
\newcommand{\Rws}{\ensuremath{R_{\rm WS}}}
\newcommand{\BR}{\ensuremath{{\cal B}}}
\newcommand{\ensuremath}[1]{\relax\ifmmode{#1}\else{$#1$}\fi}

\begin{document}        

\baselineskip 14pt
\title{$\Dz$ mixing at CLEO-II}
\author{Jeff Gronberg}
\address{UC Santa Barbara}
%
\maketitle              

\begin{abstract}        
The CLEO collaboration reported the observation of the ``wrong sign''
decay \wsdecay\ in 1993~\cite{paper_liu}.  Recent upgrades to the CLEO 
detector~\cite{CLEO}, including the installation
of a silicon vertex detector~\cite{SVX}, have increased the
sensitivity of the detector to \wsdecay.
Using \nfb\ of data we report a preliminary measurement of
the rate of the wrong sign decay \wsdecay, to be
$\BR(\wsdecay)/\BR(\rsdecay) = (\prate) $\%.
Additionally, combining this measurement with \Dz\ proper lifetime information~\cite{D_life} 
we set a limit on \DDb\ mixing of
$\BR(\mdecay)/\BR(\rsdecay) <$ \pmixinglim\ @ 90\% confidence limit.
\
\end{abstract}   	
\epsfclipon 

\section{Introduction}

Ground state mesons such as the $\KZ$, $\DZ$, and $\BZ$,
which are electrically neutral and contain a quark and
antiquark of different flavor, can evolve into their respective
antiparticles, the $\KZB$, $\DZB$, and $\BZB$.
Measurements of the rates of $\KZ\!-\!\KZB$ mixing and
$\BZ\!-\!\BZB$ mixing have guided both the elucidation
of the structure of the Standard Model, and the determination
of the parameters that populate it.  The mixing measurements
permitted crude, but accurate, estimates of the 
masses of the then-hypothetical
charm and top quark masses prior to direct observation of those
quarks at the high energy frontier.  Since \DDb\ mixing is expected
to be small in the Standard Model it provides sensitivity to
new heavy non-Standard Model particles.

The mixing of $\Dz \leftrightarrow \Dzb$ can proceed either through real 
or virtual intermediate states.
Real intermediate states ($K\pi$, $\pi\pi$, $KK$, ...) lead to lifetime
differences between the \Dz\ eigenstates (\DG\ mixing), while  
virtual intermediate states lead to a mass difference (\DM\ mixing).
Both of these amplitudes are (at least) doubly Cabibbo suppressed
compared to the total \Dz\ decay amplitude.  Additionally, the near
degeneracy of the $d$ and $s$ quark masses compared to the $W$ mass,
causes the Glashow-Illiopolouos-Maini (GIM) cancellation to be
particularly effective~\cite{DaKu}.  This introduces an additional suppression to the
mixing amplitudes of from $10$ to $10^3$.  The ratio of the
mixing amplitudes to the total decay amplitude is given by $x$ and $y$, for
virtual and real amplitudes respectively.

The observation of a value of $|x|$ in the $\DZ\!-\!\DZB$ system in 
excess of about $5\times 10^{-3}$ might  be evidence
of incomplete GIM-type cancellations among new families of particles,
such as supersymmetric partners of quarks.\cite{Coea}  The evidence would
be most compelling if either the mixing amplitude exhibited a large CP
violation, or if the Standard Model contributions could be decisively
determined.  It is possible that in the Standard Model that
$|y|>|x|$,\cite{GoPe} and a determination of $y$ allows the estimation of
at least some of the long-distance Standard Model contributions
to $x$.

\section{Formalism}

\begin{figure}[t]
\centerline{
a) \epsfxsize .4\textwidth \epsfbox[118 66 545 685]{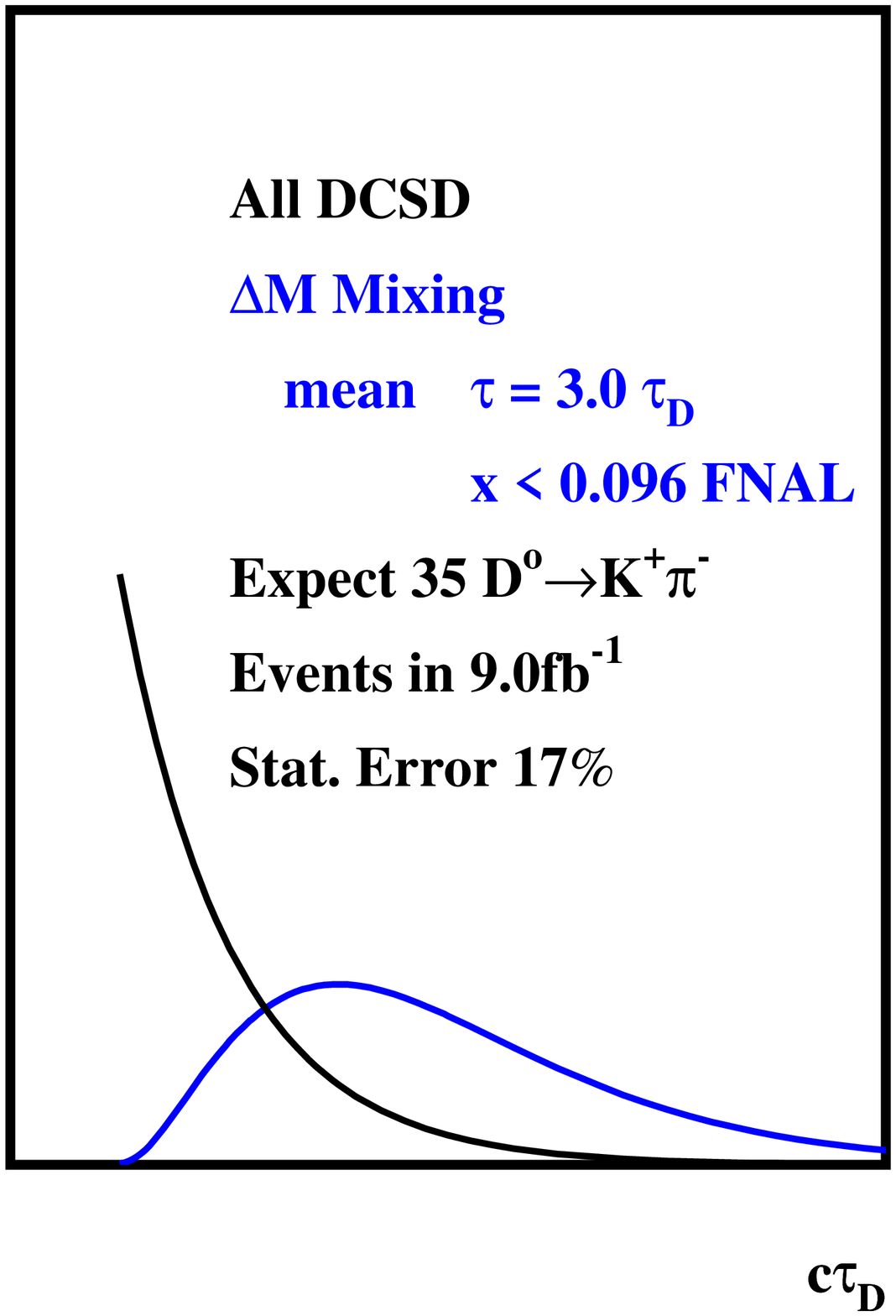} 
\hfill
b) \epsfxsize .4\textwidth \epsfbox[118 66 545 685]{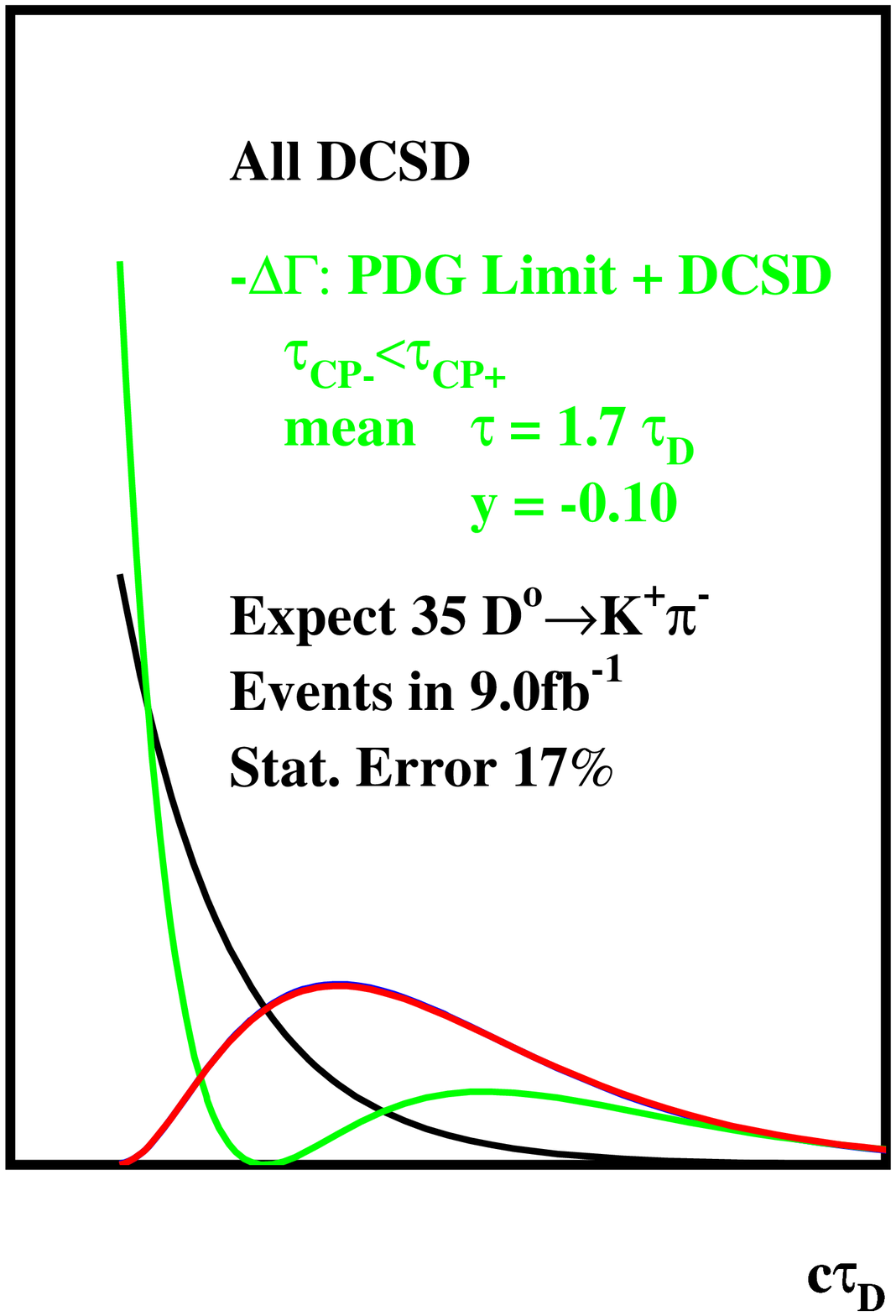}}   
\caption[]{
\label{fig:taudist}
\small Sample decay time distributions.  a) Pure DCSD (black) and \DM\ mixing (blue).  
b) Pure DCSD (black) and an interference between DCSD and \DG\ mixing (red ). }
\end{figure}

In the two-body hadronic decay \nsdecay, ``Right sign'' events are 
produced by the Cabibbo favored decay \rsdecay.
``Wrong sign'' events,
\wsdecay, are produced either through the doubly Cabibbo suppressed
tree diagram (DCSD) or through \Dz\ mixing followed by the Cabibbo allowed
decay, \mdecay.
The decay time distribution of the \Dz\ mesons allows us to
separate the total wrong sign rate into its \DG, \DM\ and
DCSD components.  In the limit of small mixing and no CP violation 
the decay time distribution
is given by~\cite{review_liu},

\begin{equation}
w(\tau)=(R_{\rm DCSD}+\sqrt{2R_{\rm DCSD}R_{\rm Mix}}\cos\phi\,\tau + \frac{1}{2}R_{\rm Mix}\tau^2)e^{-\tau}
\label{eq:taudist}
\end{equation}

where
\begin{equation}
\Rdcsd = \frac{\BR(\wsdcsd)}{\BR(\rsdecay)} = \tan^4 \theta_c,\ \ 
\Rmix = \frac{\BR(\mdecay)}{\BR(\rsdecay)} = \frac{1}{2}(x^2 + y^2),
\end{equation}
\begin{equation}
{\rm and}\ \ \phi = \arctan(-2\frac{\DM}{\DG}) + \delta_s = \arctan(-\frac{x}{y}) + \delta_s,
\end{equation}
and $\delta_s$ is the strong phase between the \wsdecay\ and \rsbdecay\ 
amplitudes and is small by theoretical bias.
Examples of the distribution for DCSD, \DM\ mixing and \DG\ mixing are plotted in 
Figure~\ref{fig:taudist}.
The time-integrated wrong-sign rate is,
\begin{equation}
\Rws=\Rdcsd+\sqrt{2\Rdcsd \Rmix}\cos\phi+\Rmix,
\label{eq:wsrat}
\end{equation}
and the mean wrong-sign decay time is,
\begin{equation}
\mtws = \frac{\Rdcsd +2\sqrt{2\Rmix \Rdcsd}\cos\phi+ 3\Rmix}
{\Rdcsd+\sqrt{2\Rmix \Rdcsd}\cos\phi+\Rmix}
\label{eq:wstau}
\end{equation}

\section{Experimental Technique}

We reconstruct the decay chain \dstdecay; \wsdecay.
The charge of the soft pion from the \Dstp\ decay 
allows us to tag the flavor of 
the \Dz\ at the moment of production.  Combining that with the 
charge of the kaon determines whether the event is 
right or wrong sign.

\begin{figure}[p]
\centerline{
a) \epsfxsize .45\textwidth \epsfbox[61 102 554 685]{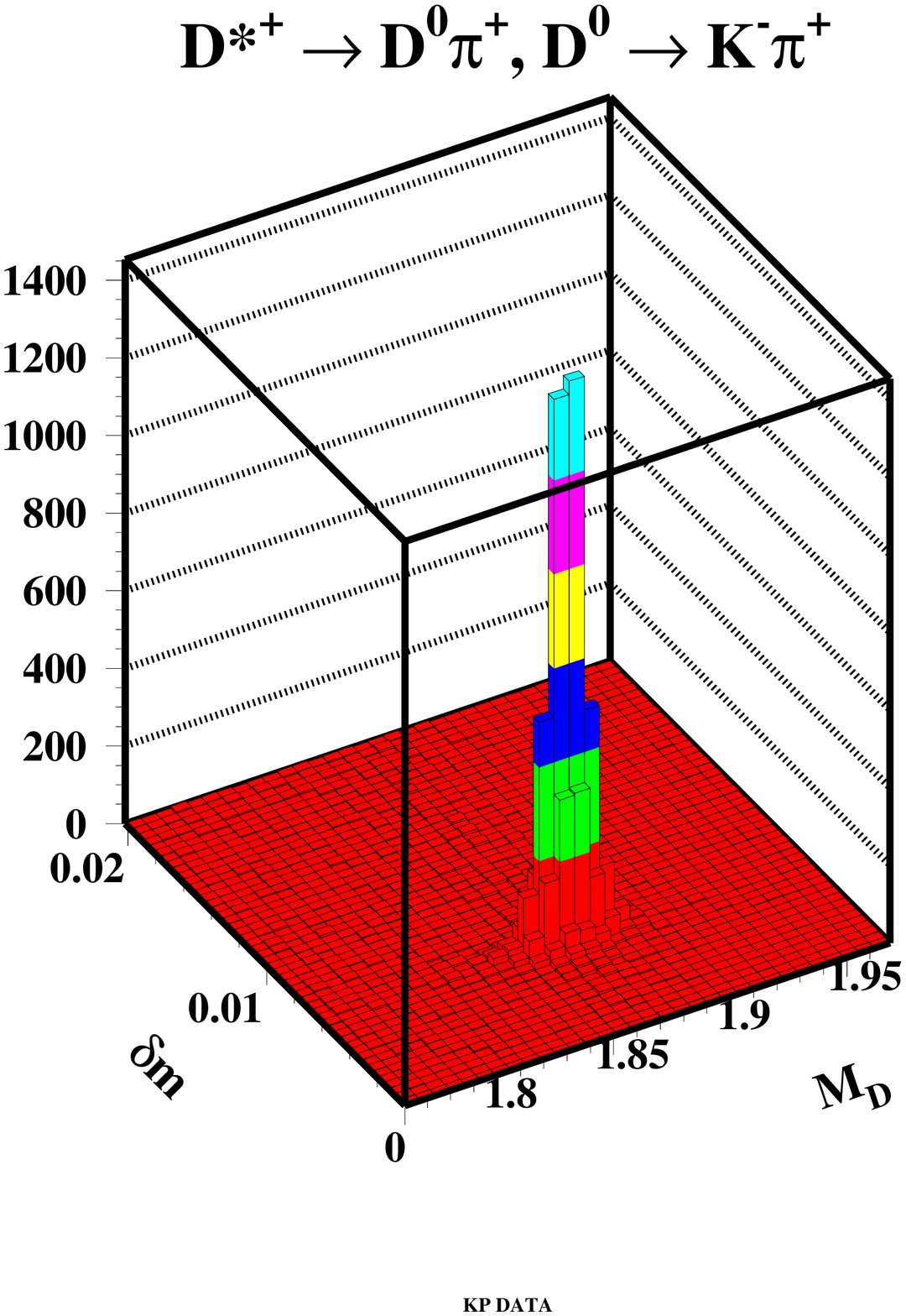} 
\hfill
b) \epsfxsize .45\textwidth \epsfbox[74 102 554 685]{gronberg0419fig3b.ps}}
\centerline{
c) \epsfxsize .45\textwidth \epsfbox[74 102 554 685]{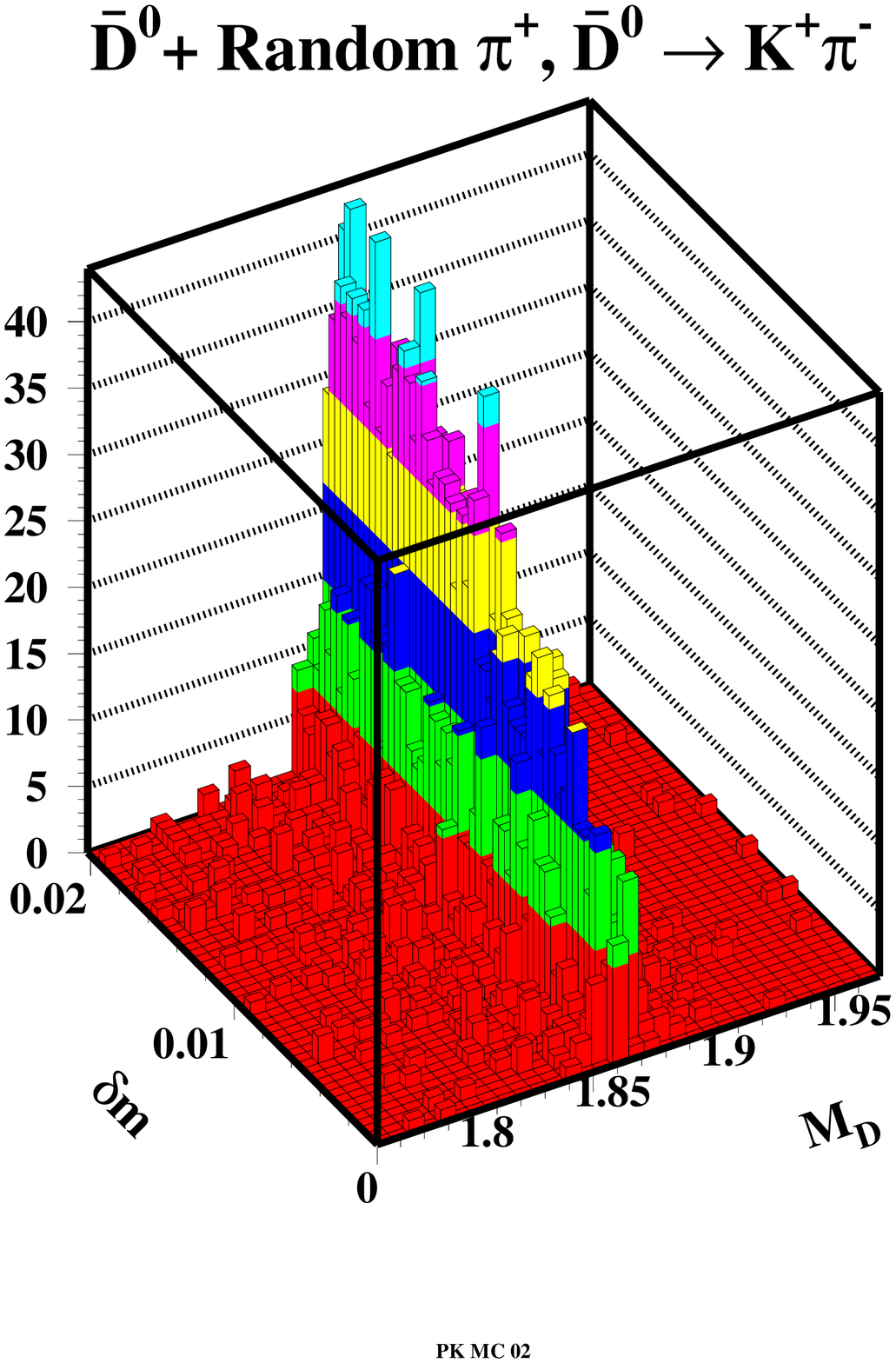} 
\hfill
d) \epsfxsize .45\textwidth \epsfbox[74 102 554 685]{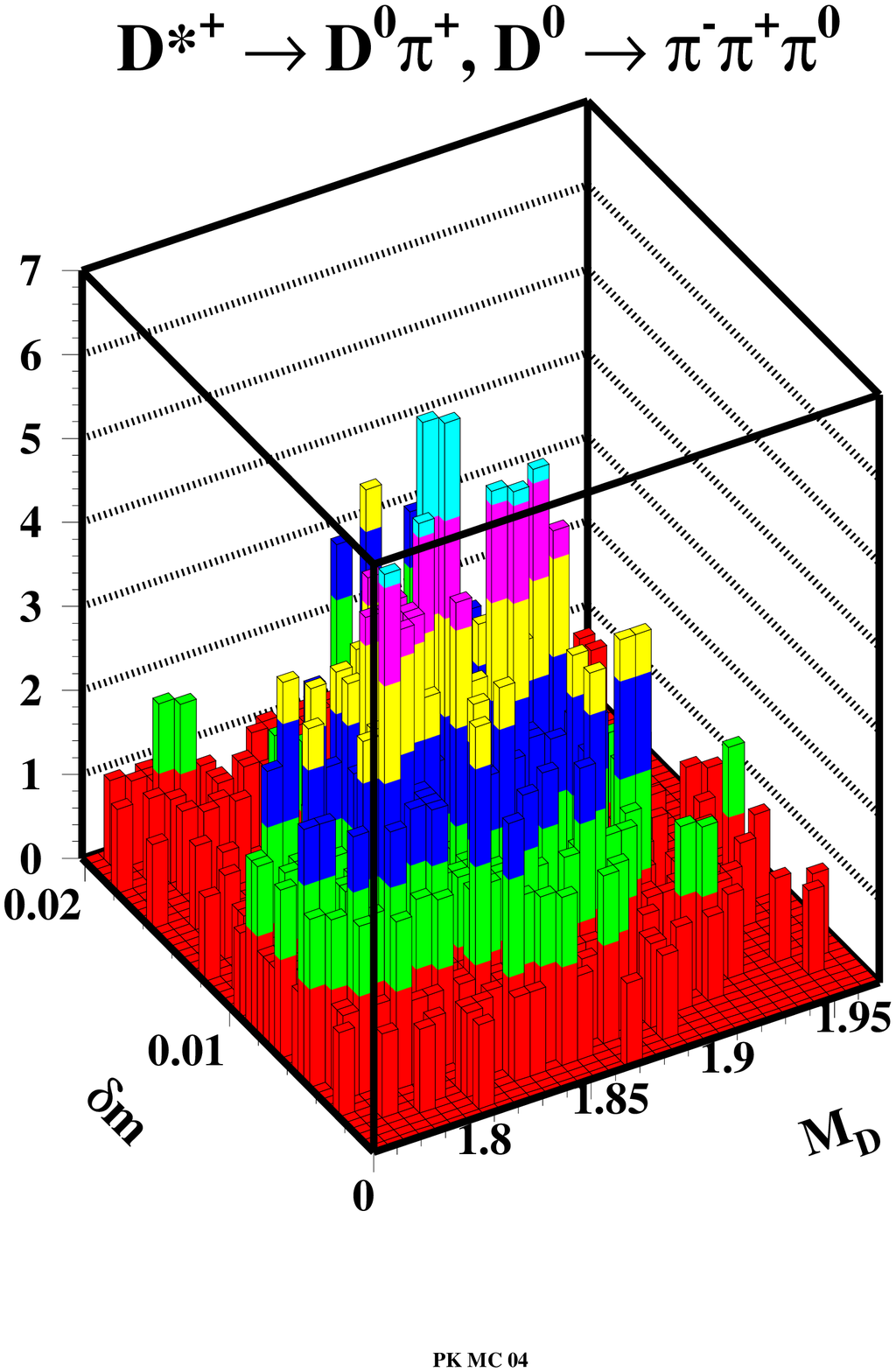}}   
\caption[]{
\label{fig:backgrounds}
\small The distributions in the \MD\ vs. \dM\ plane for; a) Right sign data events, b)
misidentified right sign Monte Carlo events, c) \rsdecay\ decays with a 
random soft track Monte Carlo events, 
d) \dstdecay\ decays with a misreconstructed \tbdecay decay Monte Carlo events.}
\end{figure}

For right sign events, as shown in
Figure~\ref{fig:backgrounds}-a, we observe the mass distribution of the $K\pi$ 
combinations (\MD) to peak at the \Dz\ mass.
Additionally, when we construct the quantity \dM, 
which is $M_{K\pi\pi}-M_{K\pi}-M_{\pi}$, we see a peak
at the \mbox{\Dst\ - \Dz\ - $\pi$} mass difference.  These events allow
us to model the wrong sign signal shape since the right sign
and wrong sign events have identical kinematics.

\begin{figure}[p]
\centerline{
a) \epsfxsize .42\textwidth \epsfbox[42 66 557 738]{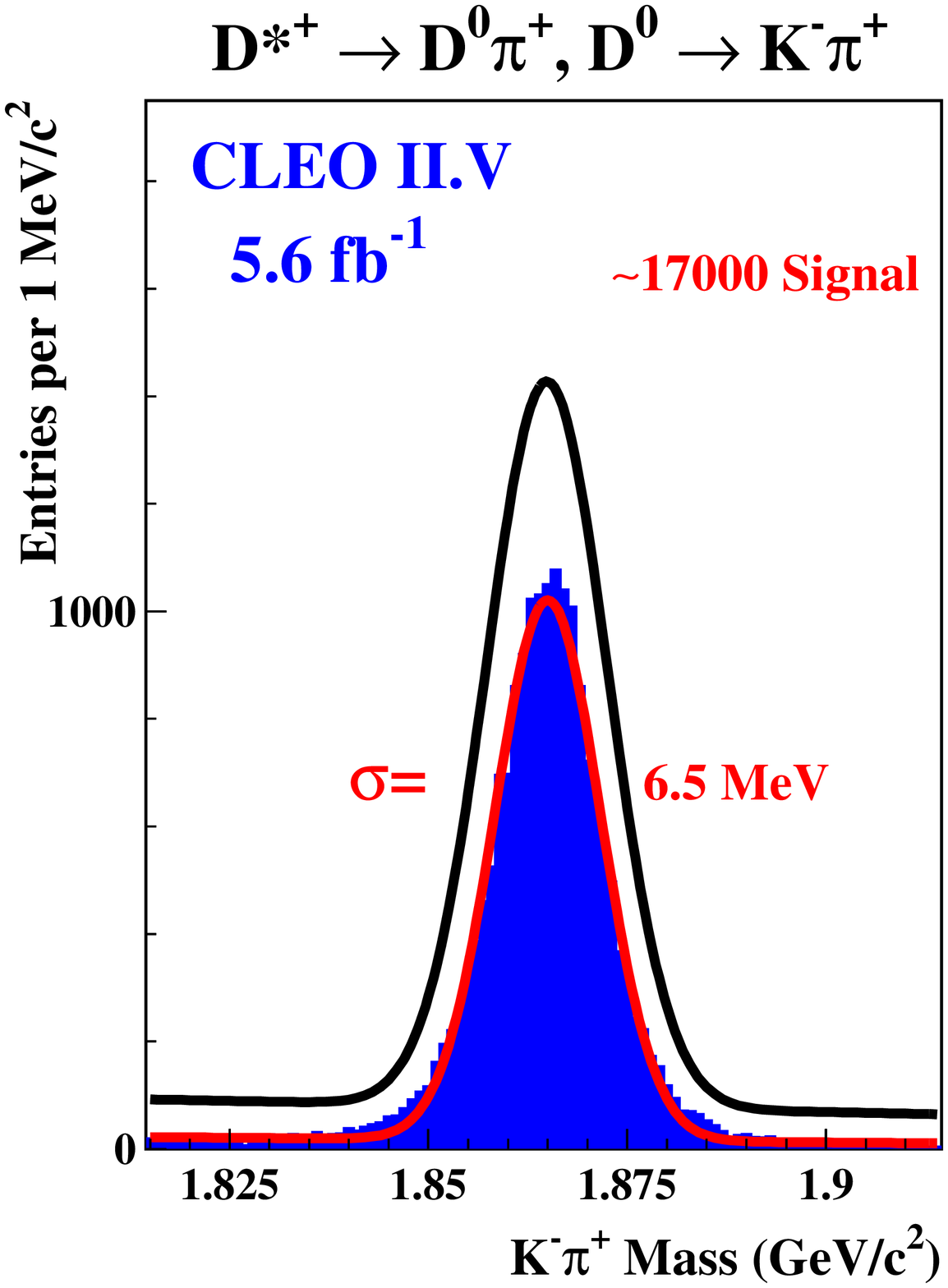} 
\hfill
b) \epsfxsize .42\textwidth \epsfbox[42 66 557 738]{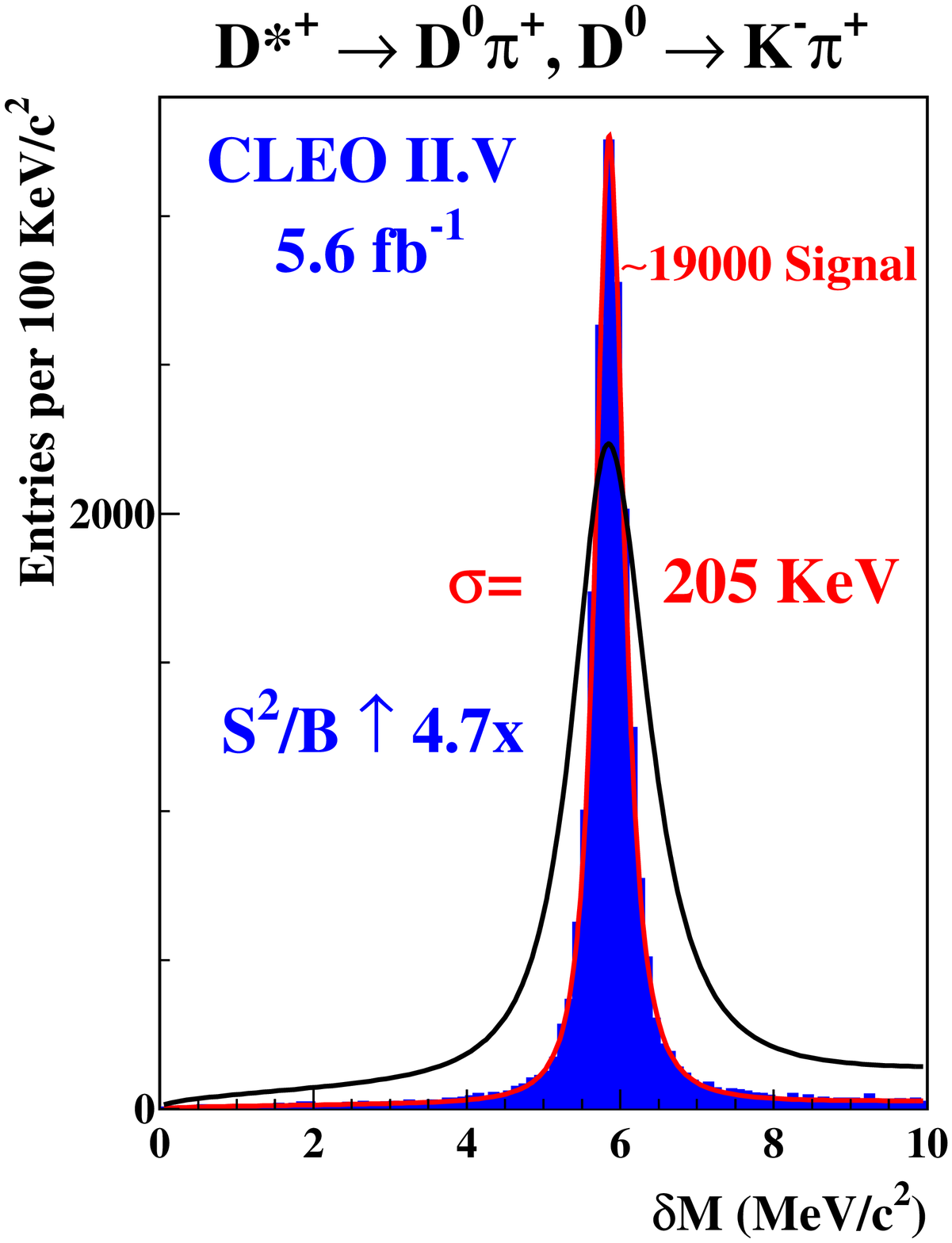}}
\caption[]{
\label{fig:resolutions}
\small The experimental resolutions in; a) \MD, b) \dM.  The black line
is the distribution before the soft pion refit.}
\end{figure}

The signal shape is Gaussian in each of the two dimensions.  The
resolution on \MD\ is 6.5 MeV and is dominated by momentum mismeasurements.
It is shown for right sign events in Figure~\ref{fig:resolutions}-a.
The resolution on \dM\ is dominated by multiple scattering of the
slow pion from the \Dst.  Using information from the silicon vertex
detector we determine the decay point of the \Dst\ by extrapolating from the
reconstructed \Dz\ spatial vertex back to the
beam spot along the \Dz\ momentum vector.
We can then refit the soft pion momentum vector 
through this additional point which improves
the measured \dM\ mass resolution from about 700 to 205 KeV.
The soft pion refit also reduces backgrounds since 
background soft pions are often inconsistent with
production at the primary vertex.  The combination of these two
effects leads to an increase in signal squared over background of 4.7,
as shown in Figure~\ref{fig:resolutions}-b.

The main backgrounds to the analysis all have shapes that are distinct
from signal in the \MD\ vs. \dM\ plane.
Figure~\ref{fig:backgrounds}b shows right sign events
where the pion from the \Dz\ decay is
misidentified as a kaon and vice versa.  This double misidentification leaves
the \dM\ peak intact but smeares out the \Dz\ peak.  
Figure~\ref{fig:backgrounds}c shows background events composed of a
properly reconstructed \rsdecay\ decay plus a random slow track.
They are indistinguishable from signal in \MD, since they are real \Dz\ decays.
However, the events distribute themselves in \dM\ according to 
phase space and produce equal numbers of right and wrong sign events.
Lastly, there are \dstdecay;$\Dz \rightarrow X$ decays.  For instance, 
if a  \tbdecay\ decay is misreconstructed it can peak broadly 
around the \Dz\ mass if the $\pi^0$ mass
is neglected and one of the other pions is misidentified as a kaon.  Since
the momentum vector of the actual and the misreconstructed \Dz\ are close,
the \dM\ distribution is broadly peaked about the 
\mbox{\Dst\ - \Dz\ - $\pi$} mass difference.

\section{Wrong-Sign rate \Rws\ and Mean Decay time \mtws}

\begin{figure}[p]
\centerline{
a) \epsfxsize .40\textwidth \epsfbox[38 64 544 732]{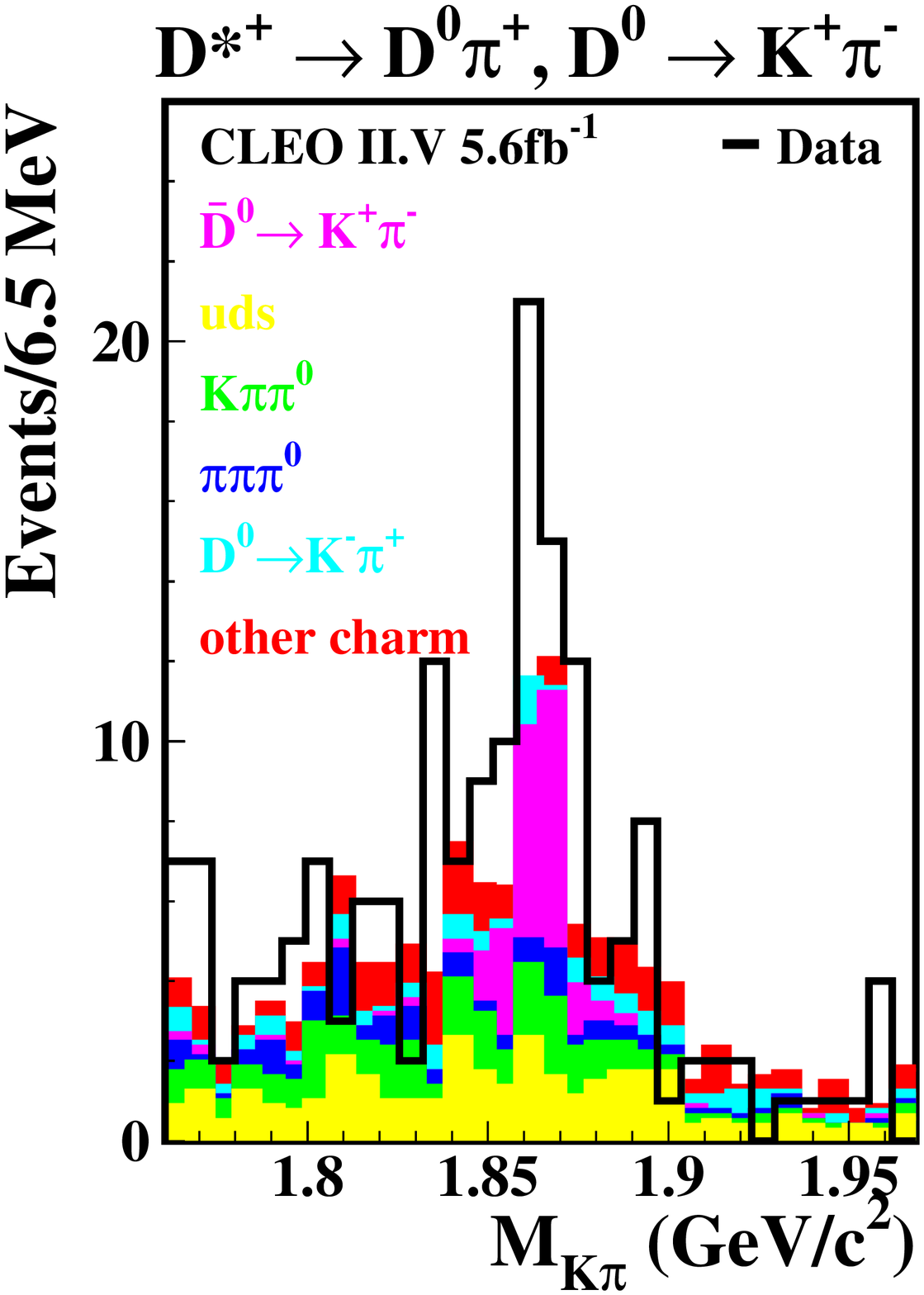} 
\hfill
b) \epsfxsize .40\textwidth \epsfbox[38 64 544 732]{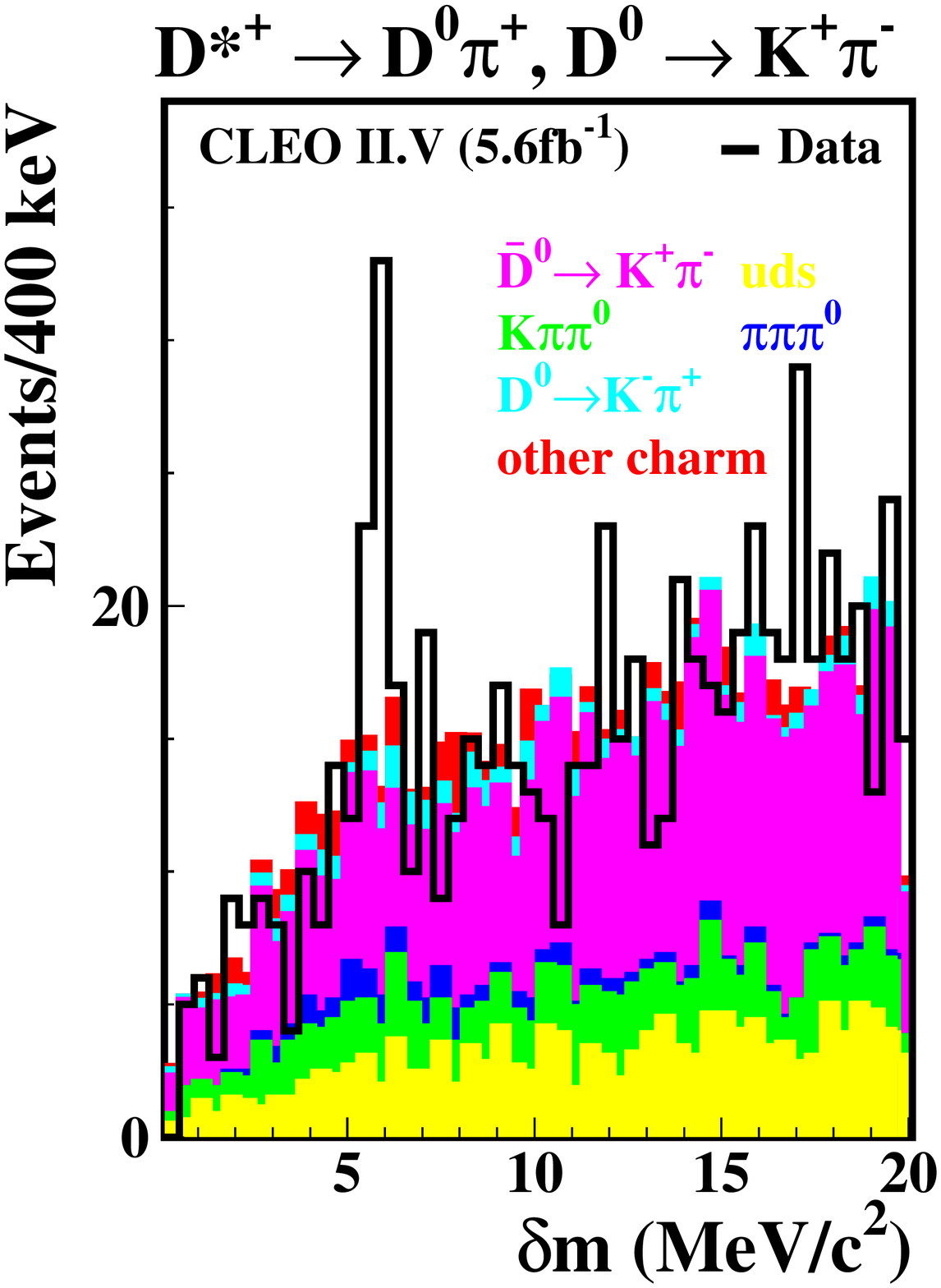}}
\caption[]{
\label{fig:wssignal}
\small Projections of the binned likelihood fit in; a) \MD, b)\dM.  The fitted
amounts of the background fractions are plotted.}
\end{figure}

We have performed a binned maximum likelihood fit of the MC-generated
background components to the two dimensional data on the
\MD\ vs. \dM\ plane, to obtain \Rws.
\begin{equation}
\Rws  = {\Gamma(\wsdecay)\over\Gamma(\rsdecay)}\,=\,0.0031 \pm .0009(stat) \pm .0007(syst)
\label{eq:rws}
\end{equation}
The fit also yields a breakdown of the background
event content in Fig.~\ref{fig:wssignal}a and ~\ref{fig:wssignal}b. 

\begin{figure}[p]
\centerline{
a) \epsfxsize .40\textwidth \epsfbox[17 53 520 765]{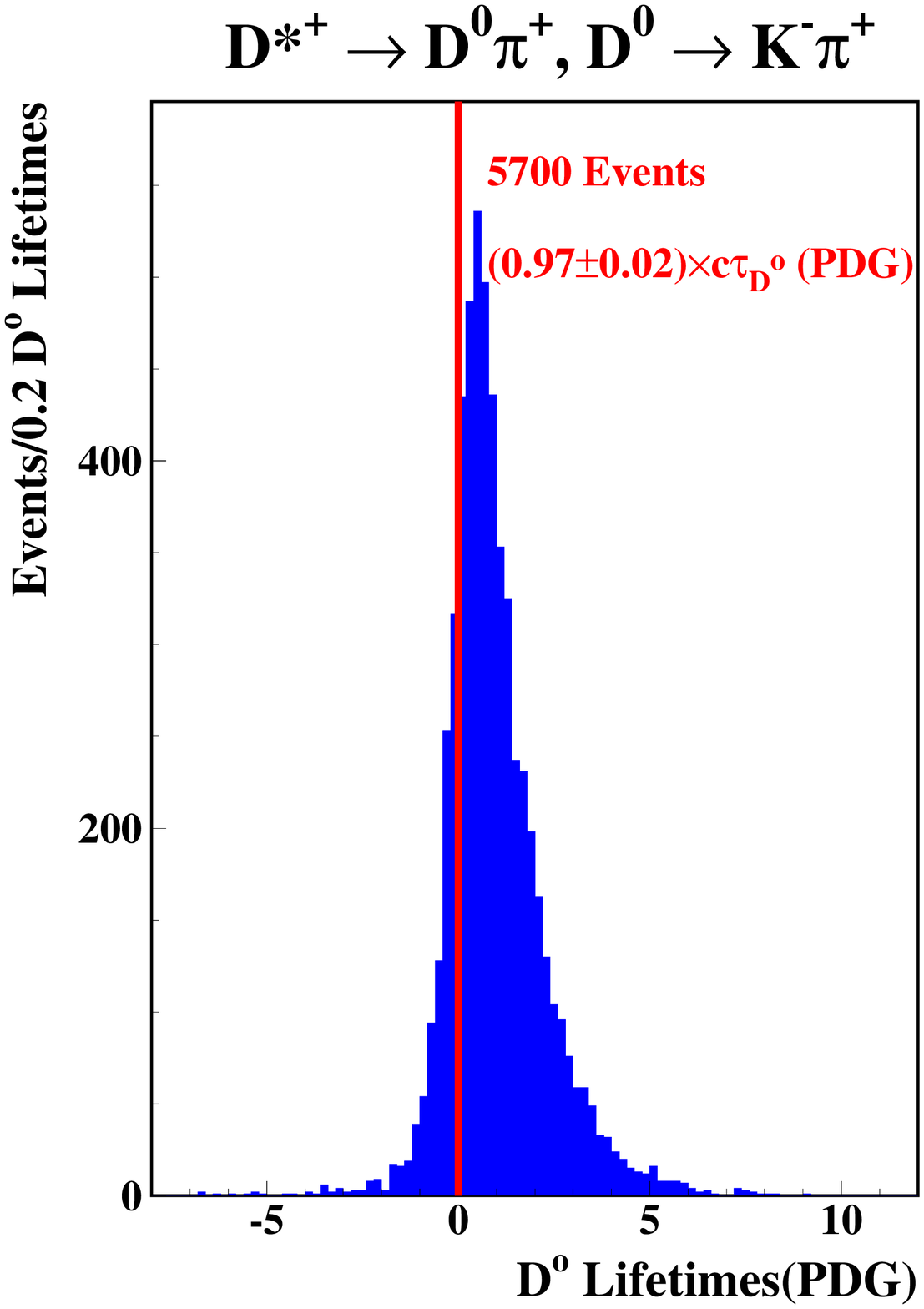} 
\hfill
b) \epsfxsize .40\textwidth \epsfbox[17 53 520 765]{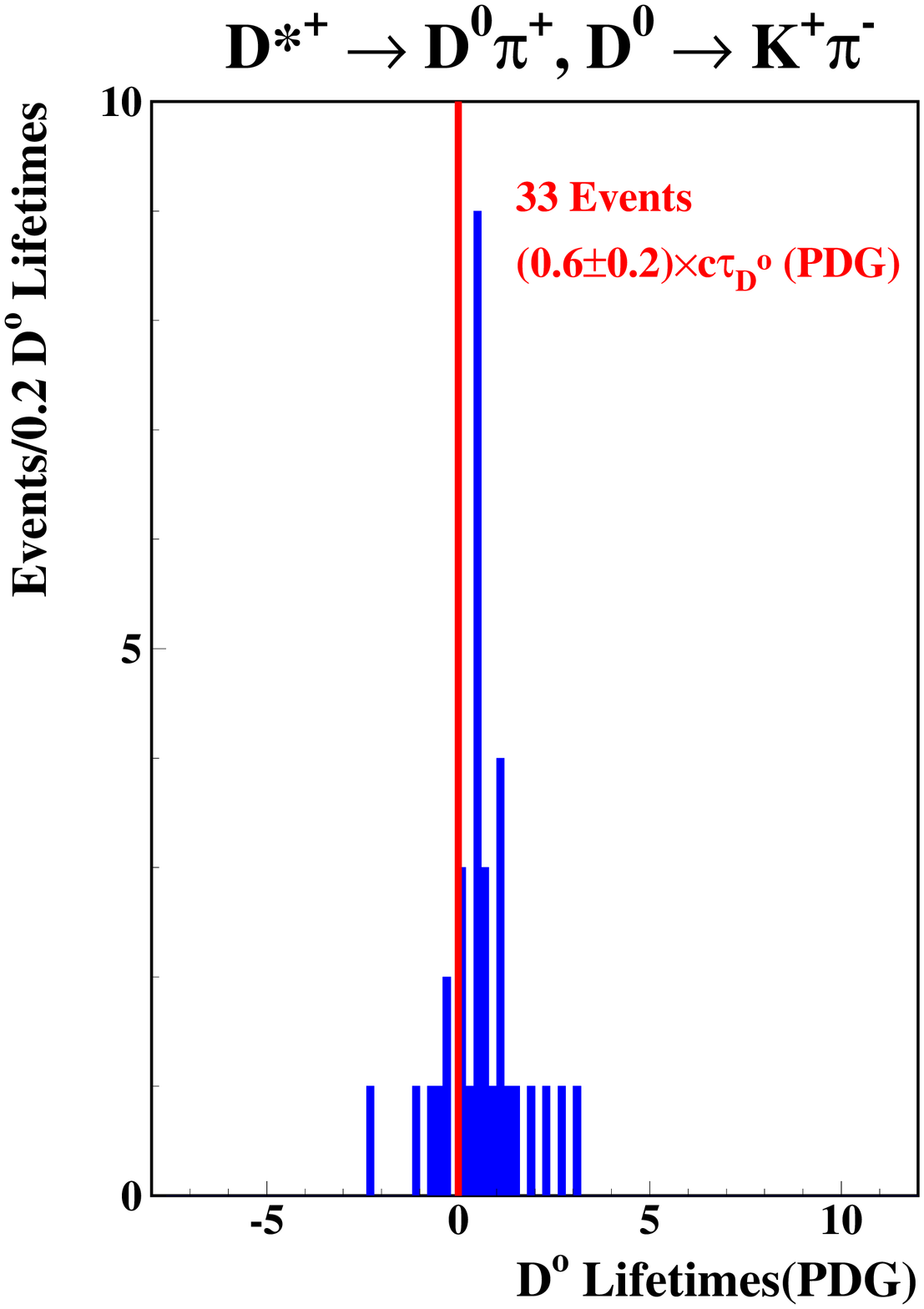}}
\caption[]{
\label{fig:lifetimes}
\small The decay time distributions for events in the signal region for; a) Right sign events,
b) Wrong sign events.}
\end{figure}

The proper time distribution of events within the signal region is shown
in Figure~\ref{fig:lifetimes} for both right sign and wrong sign
events.  Events with poorly measured proper times are excluded.
The mean proper time of the right
sign decays is $(\rslife) \times \dzct$ which is consitent with 
a pure exponential decay with the \Dz\ lifetime.  The wrong sign
events have a mean proper time of $(\wslife) \times \dzct$.  According
to the fit, the wrong sign events are composed of signal (\sigfrac),
real \Dz\ decays (\dzfrac) and other backgrounds (\othfrac).
If we assume that \Dz\ decays have a lifetime of one \dzct and the other
backgrounds have a lifetime of zero, which is a conservative assumption, 
then we determine the signal lifetime to be $(\siglife) \times \dzct$.
Care must be taken when using this measurement to determine a 90\% 
confidence level upper limit since the physically allowed region of \mtws\ ranges
between one and three for all DCSD or all \DDb\ mixing when $\cos\phi = 0$.

\section{Previous \DZ-\DZB \ Mixing Limits}
\begin{figure}[p]
\centerline{
\epsfxsize \textwidth \epsfbox{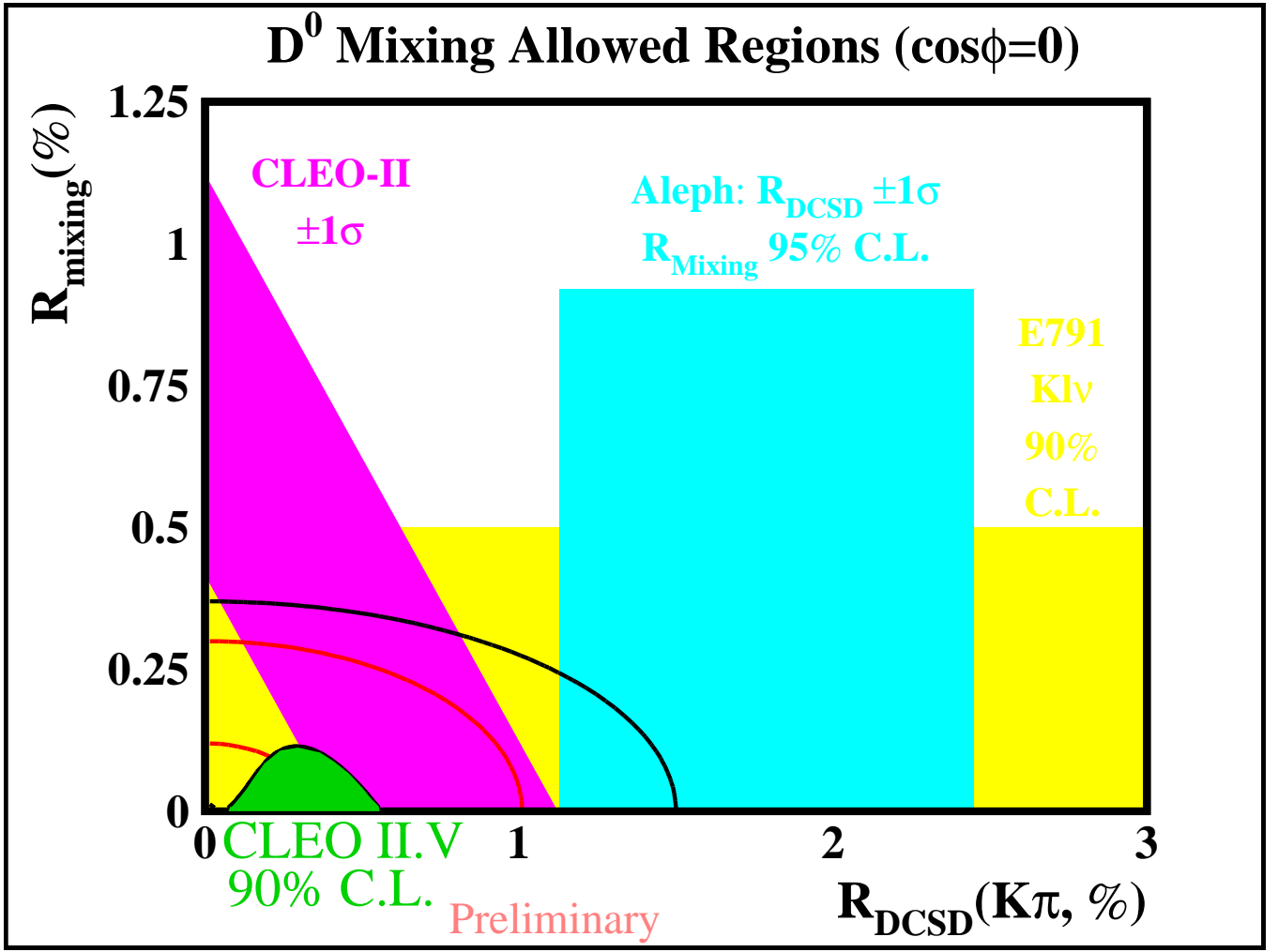}}
\caption[]{
\label{fig:world}
\small The world mixing and DCSD limits assuming $\cos\phi = 0$.}
\end{figure}

Three groups have reported non-zero measurements of $R_{\rm WS}$
evaluated for the case $\cos\phi=0$ and assuming no CP violation.
\begin{itemize}
\item CLEO-II\cite{paper_liu}, equivalent to 
      $R_{\rm WS}=R_{\rm DCSD}+R_{\rm Mix}=(0.77\pm0.35)\%$
\item E791\cite{mix_e791_had}, where $R_{\rm DCSD}=(0.68\pm0.35)\%$,
      and $R_{\rm Mix}=(0.21\pm0.09)\%$, where, for $R_{\rm Mix}$,
      $\DZ\to K^+\pi^-\pi^+\pi^-$ contribute in addition to
      $\DZ\to K^+\pi^-$; no report of a non-zero
      $R_{\rm Mix}$ was made.
\item Aleph\cite{mix_aleph}, where $R_{\rm DCSD}=(1.84\pm0.68)\%$, and
      an upper limit of $R_{\rm Mix}<0.92\%$ is obtained, at
      $95\%$ C.L.
\end{itemize}
Additionally, 
the E691 collaboration\cite{mix_e691} limited $R_{\rm Mix}<0.37\%$,
at 90\% C.L., where again $\DZ\to K^+\pi^-\pi^+\pi^-$ 
contribute in addition to $\DZ\to K^+\pi^-$,
and $R_{\rm DCSD}<1.5\%$ at 90\% C.L.
The E791\cite{mix_e791_lep} collaboration sought 
$\DZ\!\to\!K^+\ell^-\overline{\nu}_{\ell}$,
and set the limit that $R_{\rm Mix}<0.5\%$.  
The regions allowed by the above work, in the $R_{\rm Mix}$ 
vs. $R_{\rm DSCD}$ plane, for
$\cos \phi=0$, are shown in Fig.~\ref{fig:world}.

\section{CLEO-II.V Charm Mixing Limits}{\label{sec:mixlim}}
The mixing limits determined from \wsdecay with 5.6$fb^{-1}$ of
CLEO-II.V data in column 1 of table~\ref{tbl:mixtab} 
are combined with \DZ$ \rightarrow CP^+$ analysis from
E791\cite{mix_e791_CP} (table~\ref{tbl:mixtab},column 2). The
CLEO-II.V sensitivity (9.1$fb^{-1}$) combining
\wsdecay,$K^+\pi^-\pi^0,K^+\pi^-\pi^+\pi^-$ and \DZ$\rightarrow CP$
analyses is listed in column 3. A factor of 2-5 (3-10) improvement in
precision is obtained over the PDG\cite{RPP98} with 5.6$fb^{-1}$
(9.1$fb^{-1}$). The
CLEO II.V limit for $x$ $\sim\tan^2\theta_{\rm Cabibbo}$, is more or less the
largest level that \DZ-\DZB\ mixing can be in the Standard Model.

\newpage

\begin{table}[h]
\caption{Comparison of Mixing Limits. column 1:current mixing limits
determined from \wsdecay with 5.6$fb^{-1}$ of
CLEO-II.V data, column 2: E791 \DZ$ \rightarrow CP^+$ analysis combined
with column 1, column 3: Projected sensitivity of Full CLEO-II.V
dataset}
\label{tbl:mixtab}
\begin{center}
\begin{tabular}{|l|c|c|c|c|} \hline
 & CLEO-II.V  & CLEO-II.V  & CLEO-II.V & RPP98 \\ 
 &            &            & (Projected) &     \\\
 & ($5.6fb^{-1}$) & +E791 & (Final) & \\ \hline
$x$& $\hphantom{-.108<}\left|x\right|<.054$ &  $\hphantom{-.042<}\left|x\right|<.054$ &  $\left|x\right|<.03$ &  $\left|x\right|<.096$\\
$y$ & $-.108<\;y\;<.027$ & $-.042<\;y\;<.027$ & $\left|y\right|<.01$ & $\left|y\right|<.10$ \\
$R_{ws}$ & $.31\pm.09\%$ &  $.31\pm.09\%$ &  $\pm.05\%$ &  $.72\pm.25\%$\\ 
$R_{Mix}$ & $<1.1\%$ &  $<0.25\%$ &  $<0.05\%$&  $<0.5\%$\\ 
$R_{DCSD}$ & $0.1<R_{DCSD}<1.1\%$ & $0.1<R_{DCSD}<1.1\%$ & - & $<4.9\%$\\ \hline
\end{tabular}
\end{center}
\end{table}


\begin{references}  

\bibitem{paper_liu} CLEO Collab., Observation of \wsdecay
Phys.Rev.Lett.72 (1994) 1406.

\bibitem{CLEO} CLEO Collab., CLEO-II Detector,
Nucl.Instru.Meth.A320 (1992) 66-113.

\bibitem{SVX} T. S. Hill, CLEO-II Silicon Vertex Detector,
Nucl.Instru.Meth.A418 (1998) 32-39.

\bibitem{D_life} CLEO Collab., Measurement of Charm Meson Lifetimes,
\texttt{http://xxx.lanl.gov/abs/hep-ex/9902011}


\bibitem{DaKu} A.~Datta and D.~Kumbhakar, Zeit.Phys.\textbf{C27}
(1985) 515.

\bibitem{Coea} A.~G.~Cohen {\it et al.} Phys.Rev.Lett.{\bf 78}
(1997) 2300.

\bibitem{GoPe} E.~Golowich and A.~Petrov, Phys.Lett.{\bf B427}
(1998) 172.

\bibitem{review_liu} T. Liu, An Overview of \DDb\ Mixing Search
Techniques,
\texttt{http://xxx.lanl.gov/abs/hep-ph/9508415}.

\bibitem{mix_e791_had} E791 Collab., Search for \DDb\ Mixing
and Doubly-Cabibbo-suppressed Decays of the \Dz\ in Hadronic Final
States,
Phys.Rev.D57 (1998) 13.

\bibitem{mix_aleph} Aleph Collab., Study of \DDb\ Mixing
and \Dz\ Doubly Cabibbo Suppressed Decays, 
Phys.Lett.B436:211-221,1998.

\bibitem{mix_e691} E691 Collab., A Study of \DDb\ Mixing,
Phys.Rev.Lett. 60 (1988) 1239.

\bibitem{mix_e791_lep} E791 Collab., A Search for \DDb\ Mixing
 2 
in Semileptonic Decay Modes,
Phys.Rev.Lett. 77 (1996) 2384-2387.

\bibitem{mix_e791_CP} DPF99 Proceedings

\bibitem{RPP98} C. Caso {\it et al.}, European Physical Journal C
{\bf 3} (1998) 1.

 \end{references}
\end{document}